\documentclass[structabstract]{aa}  

\usepackage{graphicx}
\usepackage{natbib}
\usepackage{txfonts}
\begin{document}
   \title{Formation of the resonant system HD~60532}

   \author{Zsolt S\'andor  
          \inst{1}  
          \and  
          Wilhelm Kley\inst{2}  
          }

   \institute{Max Planck Research Group at the Max-Planck-Institut f\"ur Astronomie,  
              K\"onigstuhl 17, D-69117 Heidelberg, Germany\\  
              \email{sandor@mpia.de}  
         \and  
              Institut f\"ur Astronomie und Astrophysik, Universit\"at T\"ubingen,  
	      Auf der Morgenstelle 10, 72076 T\"ubingen, Germany\\  
             \email{wilhelm.kley@uni-tuebingen.de}  
             }

   \date{Received ; accepted }

% \abstract{}{}{}{}{}   
% 5 {} token are mandatory  
 
  \abstract 
  % context heading (optional)  
  % {} leave it empty if necessary    
   {Among multi-planet planetary systems there are a large fraction of resonant systems. 
   Studying the dynamics and formation of these systems can  
   provide valuable informations on processes taking place in protoplanetary  
   disks where the planets are thought have been formed. The recently discovered resonant  
   system HD~60532 is the only confirmed case, in which the central star hosts   
   a pair of giant planets in 3:1 mean motion resonance.}  
  % aims heading (mandatory)  
   {We intend to provide a physical scenario for the formation of HD~60532,  
    which is consistent with the orbital solutions derived from the radial velocity measurements.  
    Observations indicate that the system is in an antisymmetric configuration, while previous  
    theroretical investigations indicate an asymmetric equilibrium state.  
    The paper aims at answering this discrepancy as well.}  
  % methods heading (mandatory)  
   {We performed two-dimensional hydrodynamical simulations of thin disks with an embedded pair of massive planets.  
   Additionally, migration and resonant capture are studied by gravitational N-body   
   simulations that apply properly parametrized non-conservative forces.}  
  % results heading (mandatory)  
   {Our simulations suggest that the capture into the 3:1 mean motion resonance takes place only for   
   higher planetary masses, thus favouring orbital solutions having relatively smaller inclination   
   ($i=20^{\circ}$). The system formed by numerical simulations qualitatively show the same behaviour   
   as HD~60532. We also find that the presence of an inner disk (between the inner planet and the star)  
   plays a very important role in determining the final configurations of resonant planetary systems.  
   Its damping effect on the inner planet's eccentricity is responsible for the observed antisymmetric  
   state of HD~60532.}  
  % conclusions heading (optional), leave it empty if necessary   
   {}  
  
   \authorrunning{S\'andor \& Kley}    
   \titlerunning{Formation of HD~60532}  
   \maketitle  
%  
%________________________________________________________________  
  
\section{Introduction}  
  
A significant fraction of multi-planet planetary systems contain a pair of giant planets  
engaged in a mean motion resonance (MMR). These planets are mainly in the   
low order 2:1 MMR (as GL~876, HD~128311, and HD~73526), but in a few systems higher order resonances  
have been suggested as well; two planets around 55~Cancri may be in 3:1 MMR,   
or HD~202206 may host a pair of planets in 5:1 MMR.   
In the recently discovered system, HD~45364, the giant planets are revolving in 3:2 MMR   
\citet{Correiaetal2009A&A}.   
  
The observed orbital solutions and the formation of the majority of   
resonant systems have been studied thoroughly by many authors. It has been shown, for instance, by   
\citet{Kleyetal2004A&A} that a sufficiently slow migration process of two giant planets embedded   
in an ambient protoplanetary accretion disk ends with either a 3:1 or 2:1 resonant configuration   
depending on the speed of migration. The formation of resonant systems in 2:1 MMR has   
been modelled exhaustively by hydrodynamical and N-body simulations as well. The system GJ~876 has been   
investigated by \citet{LeeandPeale2002ApJ}, Kley et al (2005), \citet{Cridaetal2008A&A}, and the systems   
HD~128311 and HD~73526 by \citet{SandorandKley2006A&A} and \citet{Sandoretal2007A&A}, respectively.   
Most recently, the formation of the system HD~45364 with planets in the 3:2 MMR has been investigated   
by \citet{Reinetal2010A&A}. Analytical studies related to the stationary solutions of the 2:1 and 3:1 MMR have   
been done by \citet{Beaugeetal2003ApJ} and \citet{Beaugeetal2006MNRAS}, for instance.   
  
Regarding the 3:1 MMR case, the presence of the planet 55~Cancri-c has already been questioned   
\citep{Naefetal2004A&A}. Additionally, recent orbital fits indicate that the   
planets 55~Cancri-c and 55~Cancri-d are not in a resonant configuration  
\citep{Fischeretal2008ApJ}. Thus, until the recent discovery of the two planets in a 3:1 MMR around   
the F-type star HD~60532 by \citet{Desortetal2008A&A}, there has been a serious lack of knowledge   
about the observed behaviour of 3:1 resonant systems. The results of the dynamical   
study performed by \citet{Desortetal2008A&A} did not prove without any doubt that the giant   
planets are in fact in a 3:1 MMR.   
   
The final confirmation of the 3:1 MMR between the giant planets in HD~60532 is given   
in a recent paper by \citet{LaskarandCorreia2009A&A}, in which two new orbital fits are   
provided, slightly improving on the previous fit of \citet{Desortetal2008A&A} with $i=90^{\circ}$.   
Through a detailed stability analysis of different orbit integrations, \citet{LaskarandCorreia2009A&A}
suggest that an inclination of $i=20^{\circ}$ is the most likely configuration for a co-planar model.
By this assumption, the planetary masses are increased by a factor of $1/\sin(i)\approx 3$ in comparison 
to \citet{Desortetal2008A&A}. In a recent paper, \citet{2009MNRAS.400.1373L} analyse the excitation of 
{\it mutual} inclination for planetary systems driven into resonance by planet-disk interaction. 
They use a damped $N$-body evolution and find that for stronger eccentricity damping, the mutual inclination 
is less excited. However, they could not place any constraints on the observed inclination of the system
HD~60532.

The very large difference in planetary masses between the two   
orbital solutions having  $i=20^{\circ}$ or $90^{\circ}$ also raises the important question of
which orbital solutions are preferred by formation   
based on the planetary migration scenario. 
In the present paper we aim at answering this question by performing fully hydrodynamical
evolution of planets embedded in the disk. Through this procedure, we obtain realistic
migration and eccentricity damping rates that will allow us to determine the most probable
final state of the system. 
%Our results clearly favour the small inclination ( $i=20^{\circ}$) solution
%of \citet{LaskarandCorreia2009A&A}.}
 
The paper is organized as follows. First, we numerically integrate the orbits of giant  
planets using the two sets of orbital solutions given by \citet{LaskarandCorreia2009A&A} as initial conditions.  
Then by considering different planetary masses, we investigate the   
possible capture into the 3:1 MMR between the giant planets. After having formed a resonant   
system, we compare the results of our simulations to the orbital behaviour of giant planets   
obtained from numerical integrations of Sect. 2. Finally, by performing gravitational   
three-body numerical integration with dissipative forces for migration, we study how an inner   
disk influences the behaviour of the system toward its stationary solutions.  
We demonstrate that the presence of the inner disk determines the final resonant configuration  
of the system.  
  
%__________________________________________________________________  
  
\section{Orbital behaviour of the system HD~60532}  
  
\begin{table}  
\caption{Orbital solutions provided by \citet{LaskarandCorreia2009A&A}}             % title of Table  
\label{orbfit}      % is used to refer this table in the text  
\centering                          % used for centering table  
\begin{tabular}{c c c c c c c}        % centered columns (7 columns)  
\hline\hline                 % inserts double horizontal lines  
Fit & Planet & Mass [M$_J$] & $a$ [AU] & $e$ & $M$ [deg] & $\omega$ [deg]\\  % table heading   
\hline                        % inserts single horizontal line  
   I               & inner & 1.0484 & 0.7597 & 0.279 & 22.33 & 352.15 \\        
     & outer & 2.4866 & 1.5822 & 0.027 & 179.4 & 136.81 \\  
\hline       
   II               & inner & 3.1548 & 0.7606 & 0.278 & 21.95  & 352.83 \\        
     & outer & 7.4634 & 1.5854 & 0.038 & 197.53 & 119.49 \\  
\hline\hline                                   %inserts single line  
\end{tabular}  
\end{table}

%-------------------------------------------------------------  
   \begin{figure}  
   \centering  
   \includegraphics[width=\columnwidth]{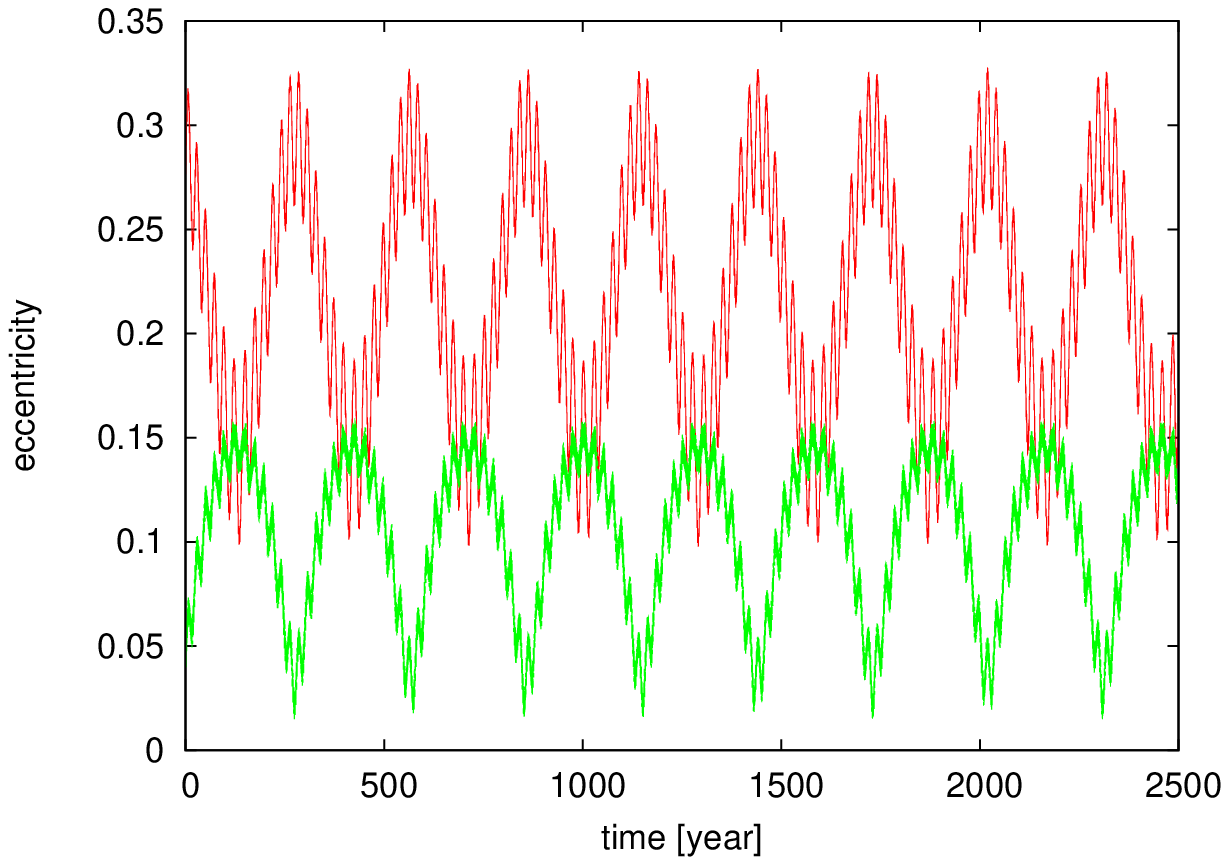}  
   \includegraphics[width=\columnwidth]{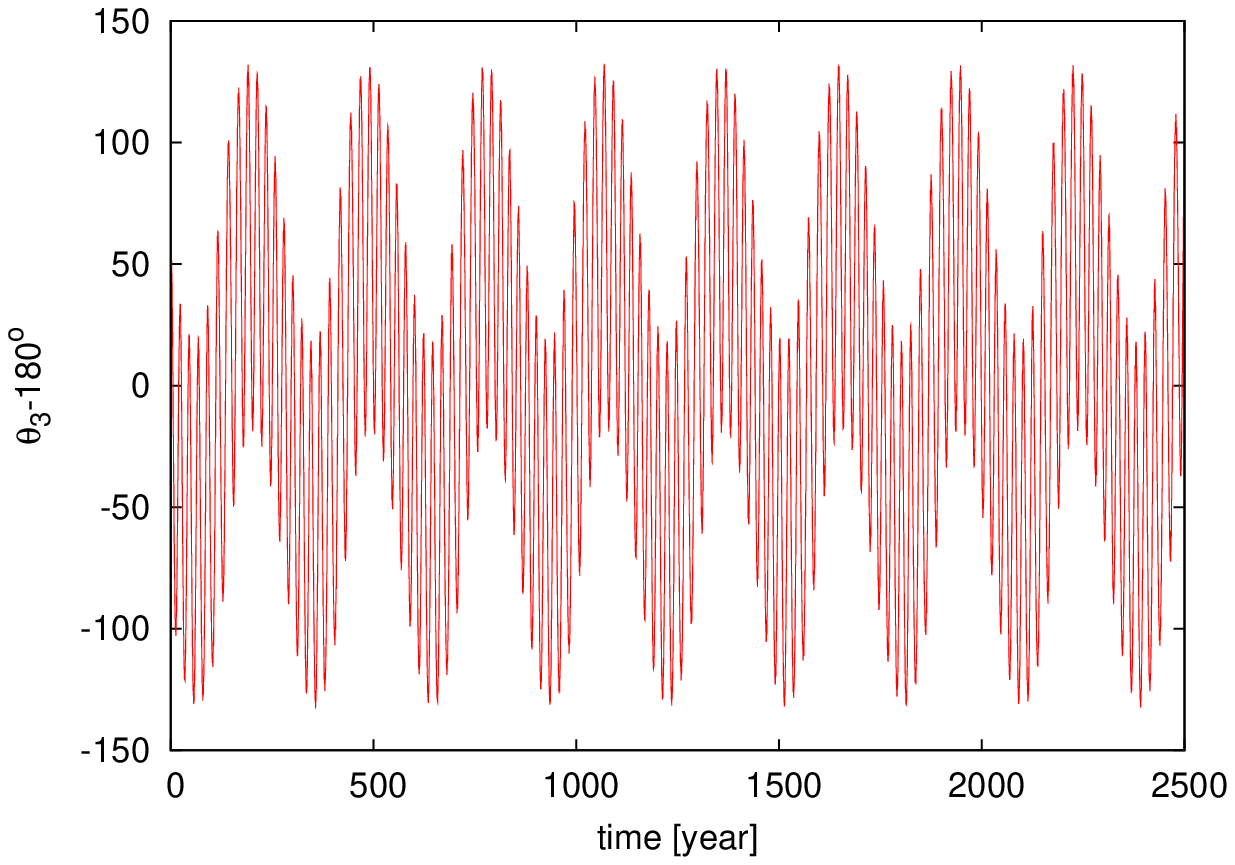}  
   \includegraphics[width=\columnwidth]{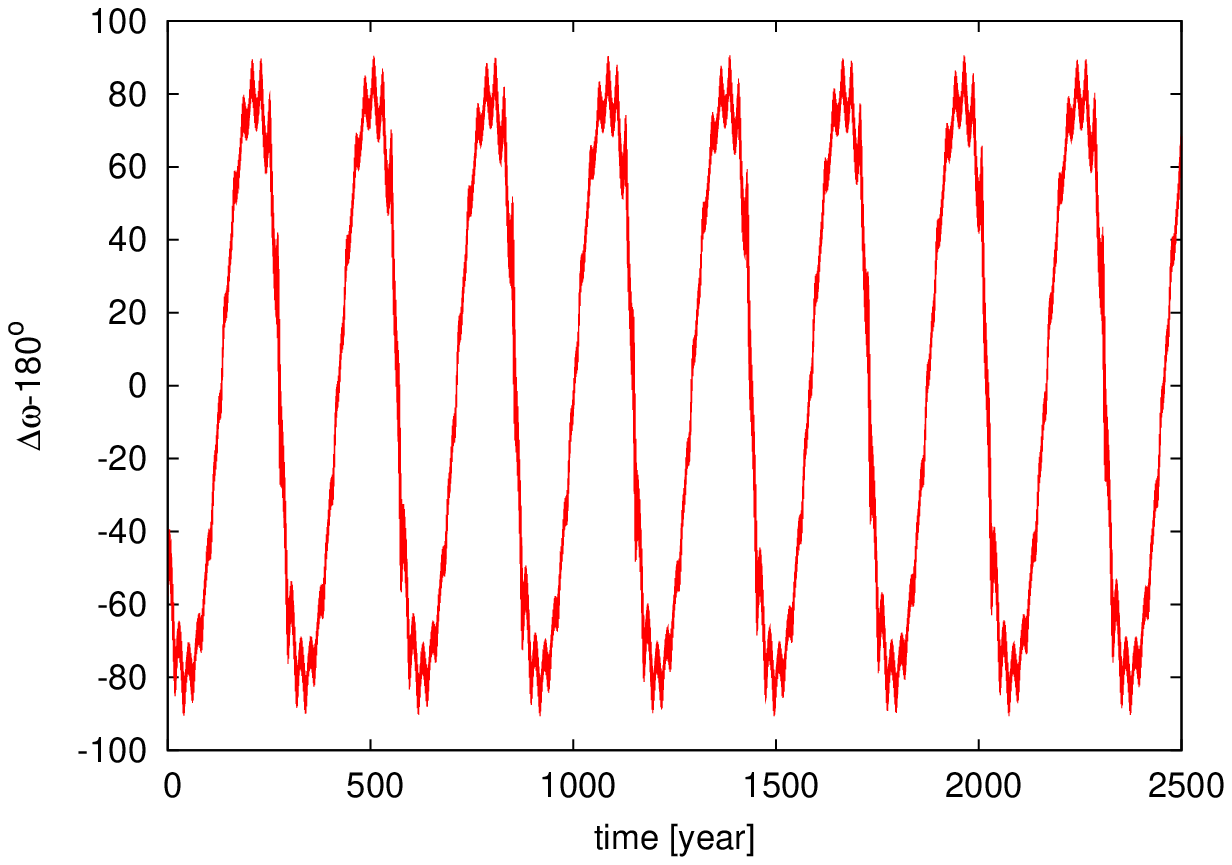}  
      \caption{Dynamical behaviour of the system HD~60532, where the initial conditions  
               of the numerical integration are calculated from Fit II of \citet{LaskarandCorreia2009A&A}.  
	       The upper panel shows the evolution of the eccentricities (inner planet red curve,  
	       outer planet green curve), middle panel the behaviour of the resonant angle $\theta_3$,  
	       while the bottom panel shows the time evolution of the corotation angle $\Delta\omega$.}  
         \label{ecc_sini90}  
   \end{figure}  
%--------------------------------------------------------------  

In this section we present the orbital behaviour of the giant planets around the star HD~60532.  
To integrate the equations of the three-body problem, we used a Bulirsch-Stoer numerical integrator.   
The initial conditions for our simulations were calculated from the orbital solutions given by   
\citet{LaskarandCorreia2009A&A}, see Fits I and II in Table \ref{orbfit}.  
  
A resonant system can be characterized by studying the behaviour of the resonant angles.    
According to \citet{MurrayandDermott1999ssd}, in the planar case of the $3:1$ MMR, there are three  
different resonant angles:  
$\theta_1 = 3\lambda_2 - \lambda_1-\omega_1-\omega_2$,     
$\theta_2 = 3\lambda_2 - \lambda_1-2\omega_2$, and
$\theta_3 = 3\lambda_2 - \lambda_1-2\omega_1$,   
where $\lambda_i$ denotes the mean longitude of the $i$-th planet, and $\omega_i$ its periastron.  
Here index ``1'' stands for the inner and index ``2'' for the outer planet.  
By introducing the corotational angle $\Delta\omega = \omega_2 - \omega_1$ and
arbitrarily choosing a $\theta_i$, all the remaining resonant angles can be expressed with $\theta_i$  
and $\Delta\omega$. If one of $\theta_i$ ($i=1,2,3$) librates around a constant value, the system   
is in a $3:1$ MMR. If $\Delta \omega$ also librates, the system is said to be in apsidal corotation.  
  
By studying Fig. \ref{ecc_sini90}, one can follow the behaviour of the eccentricities and resonant angles as   
functions of time for Fit II. The eccentricities show quite large oscillations, the inner   
planet's eccentricity varies between $e_1 \sim 0.1 - 0.33$, while the outer planet's eccentricity   
is between $e_2 \sim 0 - 0.15$. All resonant angles $\theta_i$ are librating, $\theta_1$  
around $0^{\circ}$, $\theta_2$, $\theta_3$, and $\Delta\omega$ around $180^{\circ}$. This means  
that the system is in 3:1 MMR and also in apsidal corotation, and furthermore the orbits   
of the giant planets are anti-aligned.   
In Fig. \ref{ecc_sini90} we display the behaviour of the resonant angle $\theta_3$,  
\citep[as used by][]{LaskarandCorreia2009A&A} and $\Delta\omega$.   
By using Fit I, we found the same dynamical behaviour of the  
giant planets as \citet{LaskarandCorreia2009A&A}, therefore we do not display these results here.  
  
Regarding the behaviour of the resonant angles and the possible apsidal corotation, a comprehensive   
analytical study of the 3:1 MMR has been done by \citet{Beaugeetal2003ApJ}. The authors find  
that the resonant angle $\theta_3$ and $\Delta\omega$ both librate around $180^{\circ}$ if   
the eccentricity of the inner planet is below a certain limit: $e_1 \approx 0.13$. Above this  
limit (i.e. $e_1>0.13$), there appears asymmetric libration of the resonant angles, whose mean values   
depend on the eccentricity of the inner planet. This result has also been confirmed by   
\citet{Kleyetal2004A&A}, where the mean value of the corotation angle was found to be   
$\Delta\omega \approx 110^{\circ}$.   
Returning to the behaviour of the system HD~60532 (obtained from radial velocity measurements),   
although the inner planet's eccentricity is higher than the limit $e_1=0.13$, the corotation angle  
does not yield any asymmetric libration; on the contrary, it librates around $180^{\circ}$, see the bottom   
panel of Fig. \ref{ecc_sini90}. The reason for this behaviour might be that, for some reason,   
the system could not reach the stationary solution found by \citet{Beaugeetal2003ApJ} during the planetary   
migration. A possible explanation of this phenomenon will be presented in Sect. 5.   
  
In what follows, we attempt to reproduce the dynamical behaviour of the system shown in Fig.   
\ref{ecc_sini90} by a migratory formation scenario using hydrodynamical and gravitational 3-body   
simulations with properly parametrized non-conservative forces for migration.

%__________________________________________________________________  
  
\section{Numerical setup of hydrodynamical simulations}  
  
To study the formation of the resonant system HD~60532 through migration of the giant planets,   
we performed a series of full hydrodynamical simulations of a protoplanetary accretion disk containing two   
embedded giant planets. For the hydrodynamical simulations we used the locally isothermal version of the {\tt FARGO}   
public hydrodynamical code \citep{Masset2000A&AS} and the code {\tt RH2D} \citep{1999MNRAS.303..696K}  
both well-suited to investigating planet-disk interactions.    
  
For our simulations, we used a flat accretion disk with an aspect ratio $H/r=0.05$ extending between   
$r_{min}=0.2$ and $r_{max}=5.0$ in dimensionless units. The disk has an initial surface density   
profile $\Sigma(r) = \Sigma_0 r^{-1/2}$, where the dimensionless surface density value at one distance unit is   
$\Sigma_0=3.1\times 10^{-4}$.   
This value corresponds to 1\% of the stellar mass ($M_*=1.44 M_{\odot}$).   
During our simulations, the accretion of the gas in the protoplanetary disk was driven by an alpha-type   
viscosity with $\alpha = 0.01$. A torque cut-off around the planet of $0.6 R_{Roche}$ was also applied.   
The computational domain (bounded by $r_{min}$ and $r_{max}$) was covered by 256 radial and 500   
azimuthal gridcells. The radial spacing is logarithmic, while the azimuthal is equidistant, resulting   
in nearly quadratic gridcells (meaning that at each radius $r\Delta\phi\approx \Delta r$, approximately).  
  
\begin{figure}  
   \centering  
   \includegraphics[width=\columnwidth]{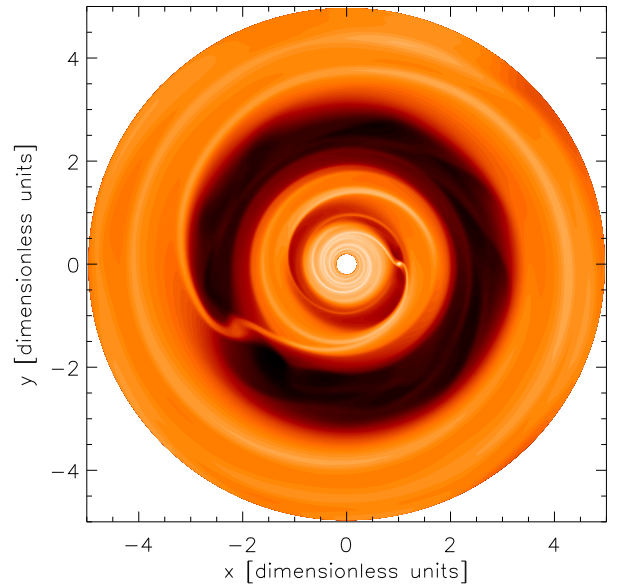}  
      \caption{The surface density distribution of the protoplanetary accretion disk after  
               500 periods of the inner giant planet. The disk model is described in Sect. 3,  
	       the planetary masses correspond to Fit II. Light colours are for higher, while  
	       dark colour shades are for low values of the surface density. For a better   
	       visualization we used a logarithmic scale.}  
         \label{disk_density}  
   \end{figure}

The planets were initially placed at $r=1.0$ and $2.5$ distance units from the star.  
To avoid the transient effects and obtain a quasi steady state initial setup,  
we first integrated the disk-planets-star system keeping the orbits of giant  
planets fixed on circular orbits until 500 orbital periods of the inner planet. The surface density   
distribution of the accretion disk obtained after the above integration time is shown in Fig.   
\ref{disk_density}, where the planetary masses correspond to the higher masses of Fit II,  
and we used a logarithmic scale for the surface density values for better visibility.   
  
There is an essential difference between our hydrodynamical setup and the one used by  
\citet{Kleyetal2004A&A} (Model C in the cited paper). In the latter model, the initial surface density profile  
tends to zero with $r\rightarrow 0$, because the inner disk was assumed to be cleared by the accretion  
of the gas to the star.   
%% (The emptying of the outer disk is prevented by the outer giant planet providing a tidal barrier   
%% to the gaseous material.)   
In such a case, the outer planet feels the negative torques of only the outer disk and migrates   
inward until it captures  the inner planet into the 3:1 MMR, which revolves in a practically gas-free   
environment. On the other hand, as demonstrated recently by \citet{Sandoretal2007A&A} and   
\citet{Cridaetal2008A&A}, the presence of the inner disk plays a very important role in damping the eccentricity of the inner   
planet during the migration process, hence in shaping the final behaviour of the system. Therefore in our   
disk model we did not allow the inner disk to be emptied, so the planets (mainly the inner one) can feel its   
effect.  
  
Having obtained almost cleared gaps after 500 orbital revolutions of the inner planet, the planets   
were released and their orbital evolution was followed until 4000 periods of the inner planet.  
For the planetary masses we used the values published by \citet{LaskarandCorreia2009A&A} (see Fits I and II),   
and also an intermediate value calculated for $i=30^\circ$.  
  
%-------------------------------------------------------------  
%--------------------------------------------------------------  
%-------------------------------------------------------------  
   \begin{figure}  
   \centering  
   \includegraphics[width=\columnwidth]{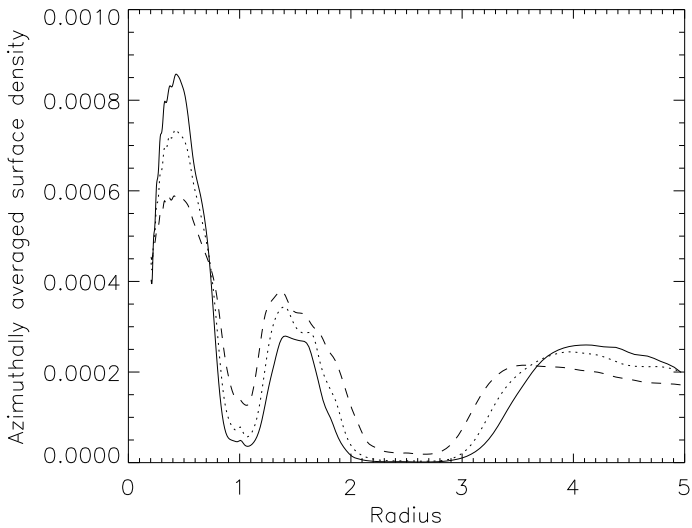}  
      \caption{The azimuthally averaged surface density profile of the protoplanetary disk after  
               500 periods of the inner giant planet for different planetary masses. Dashed line  
	       corresponds to planetary masses of Fit I, dotted line for the intermediate values,  
	       which are twice the masses of Fit I, and solid line for masses of Fit II.}  
         \label{avsurf_density}  
   \end{figure}  
%--------------------------------------------------------------  

\section{Resonant capture for different masses of the giant planets}  
  
In this section we show the results of our hydrodynamical simulations obtained by using different   
planetary masses.   
  
First we investigated the behaviour of the giant planets having low masses, corresponding  
to the Fit I, where $i = 90^{\circ}$. In this case $m_1=1.048 M_J$, $m_2=2.487 M_J$ corresponding   
to $q_1=0.0007$ and $q_2=0.00165$ dimensionless mass units (the stellar mass is unity).   
We used index ``1'' for the inner and index ``2'' for the outer planet.  
After releasing them, however, the planets did not show a convergent migration, and no   
resonant capture took place. This was also true for the intermediate value of the  
inclination  $i=30^{\circ}$, where the planetary masses are a factor of two higher than in Fit I.  
%%The reason for this lack of capture is the too rapid inward migration of the inner planet, which  
%%evades capture (see below).  
  
Finally, we used planetary masses corresponding to Fit II, where $i=20^{\circ}$, and the planetary   
masses are $m_1 = 3.15 M_J$ and $m_2 = 7.46 M_J$ or, in dimensionless mass units, $q_1=0.0021$ and   
$q_2=0.005$. (These values are quite close to the planetary masses used by Kley et al. 2004.)  
In this case the giant planets also begin to migrate, but contrary to the previous two cases,   
the outer planet migrates faster than the inner one resulting in a resonant capture.   

%-------------------------------------------------------------  
   \begin{figure}  
   \centering  
   \includegraphics[width=\columnwidth]{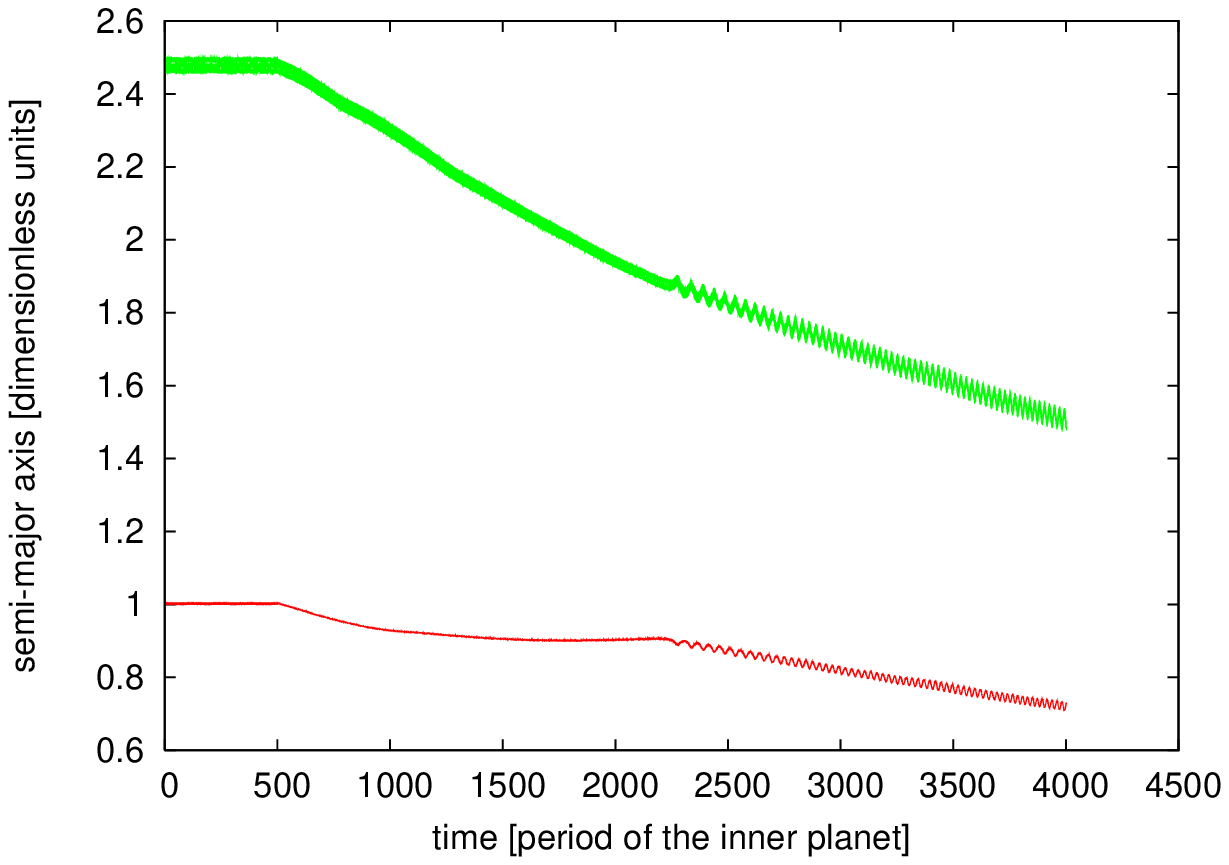}  
   \includegraphics[width=\columnwidth]{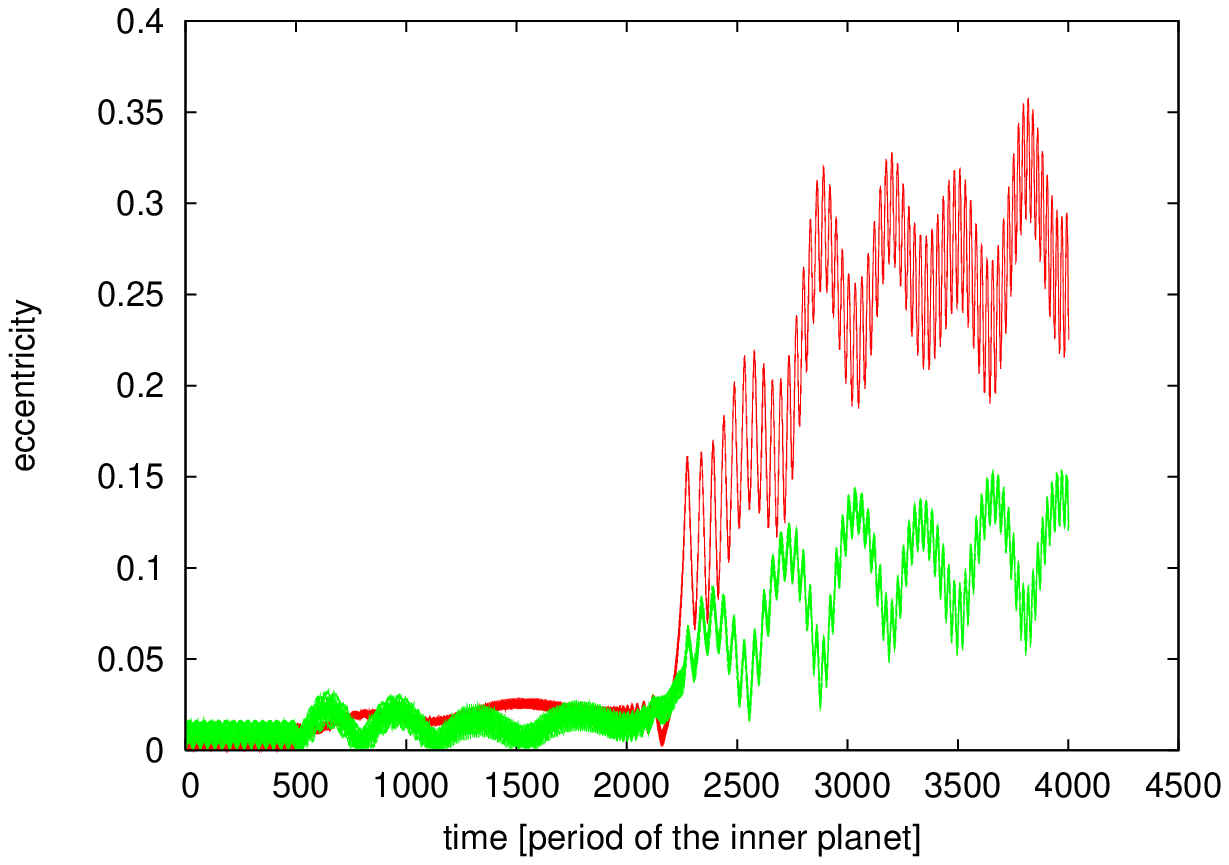}  
      \caption{Dynamical behaviour of the two giant planets embedded in the surrounding  
               protoplanetary disk: semi-major axes and eccentricities.  
	       During the first 500 orbital periods (of the inner planet), the planets are kept   
	       fixed to obtain a steady state in the disk.  
               The upper panel shows the behaviour of the semi-major axes, while the bottom panel  
	       shows the time evolution of the eccentricities (red line corresponds to the   
	       inner, green line to the outer giant planet). The capture into the 3:1 MMR occurs  
	       around 2300 periods of the inner planet. The planetary masses   
	       ($m_1 = 3.15 M_J$ and $m_2 = 7.46 M_J$) are taken from Fit II.}  
         \label{formation_sma_ecc}  
   \end{figure}  
%--------------------------------------------------------------  
  
%-------------------------------------------------------------  
   \begin{figure}  
   \centering  
   \includegraphics[width=\columnwidth]{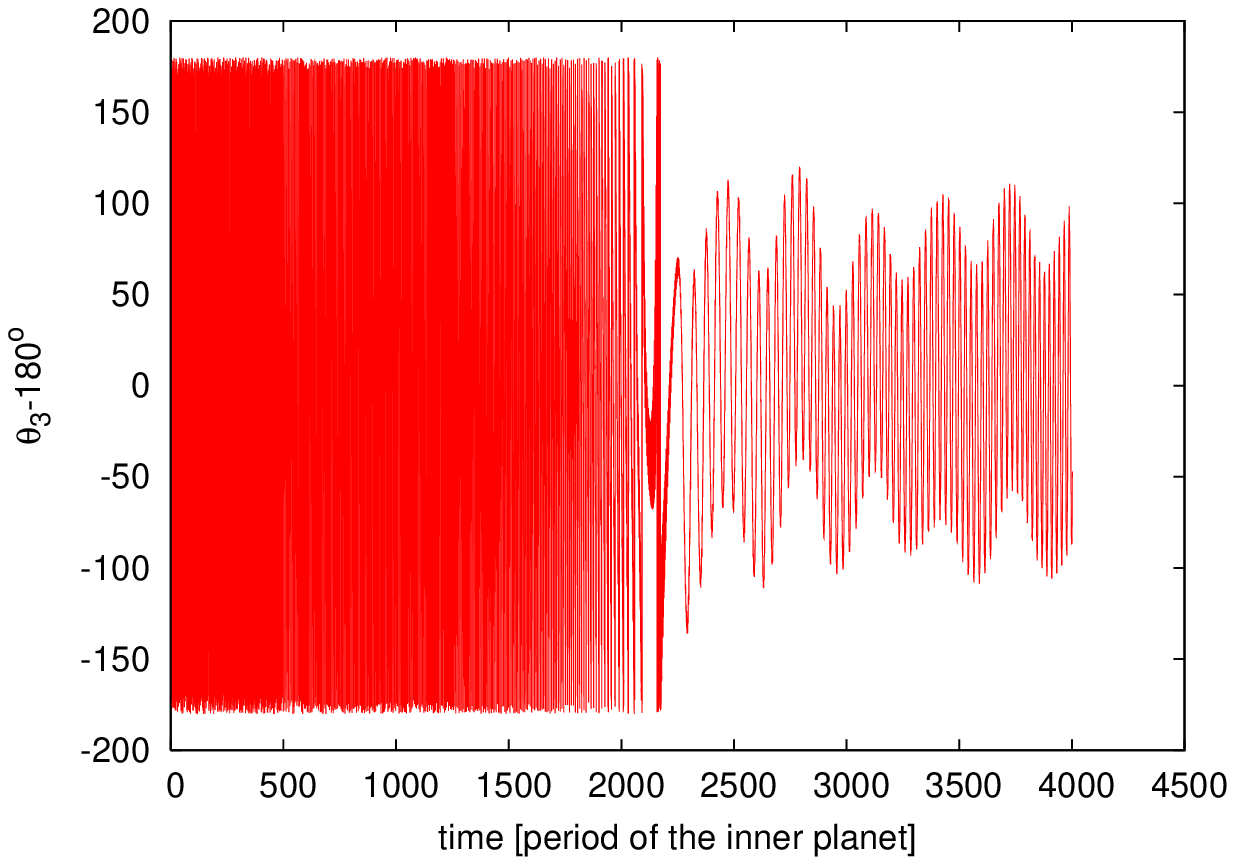}  
   \includegraphics[width=\columnwidth]{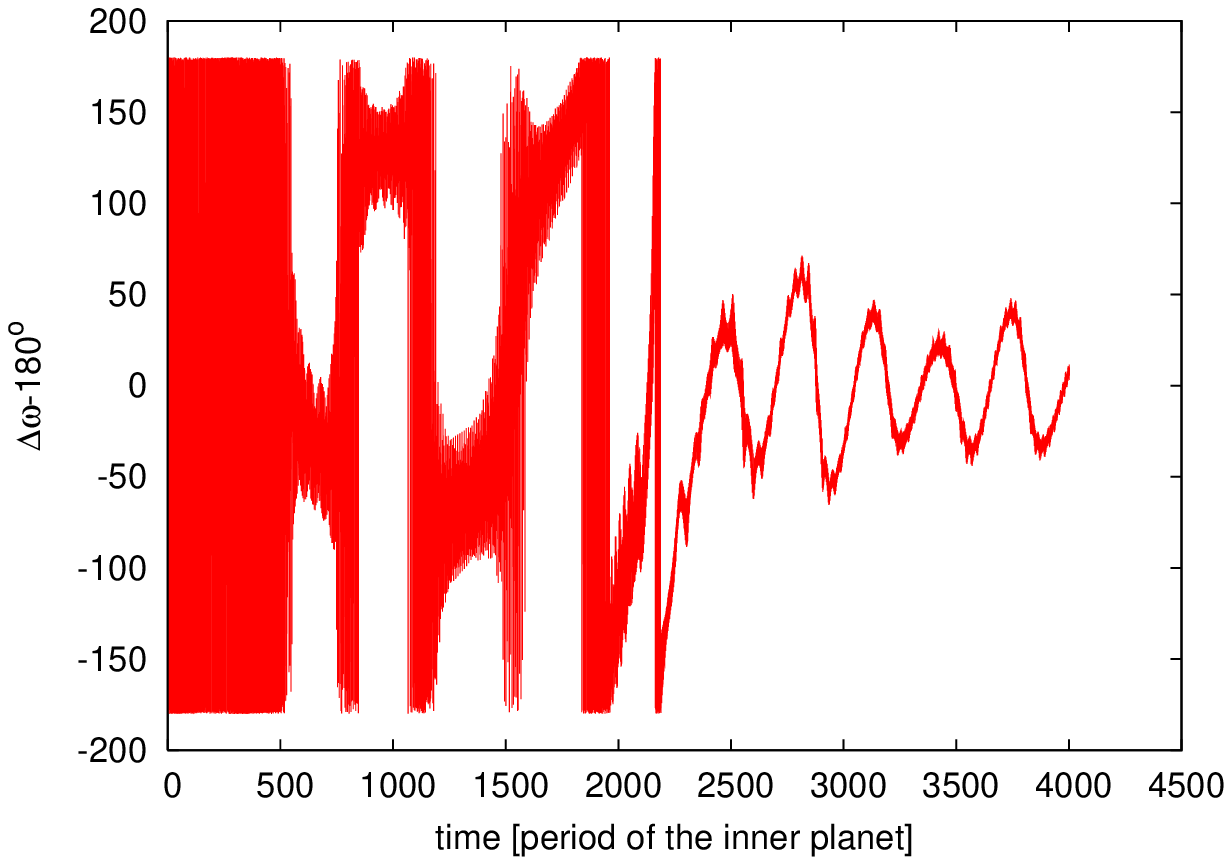}  
      \caption{Dynamical behaviour of the two giant planets embedded in the surrounding  
               protoplanetary disk for Fit II.  
               The evolution of the resonant angle $\theta_3$ and $\Delta\omega$   
               are shown in the top and bottom panels, respectively.   
	       After entering into the 3:1 MMR, the orbits of the giant   
	       planets exhibit apsidal corotation.}  
         \label{formation_resang}  
   \end{figure}  
%--------------------------------------------------------------
  
Before describing the dynamical behaviour of the system after the capture into the resonance, we should   
comment on why this phenomenon does not occur in the cases of lower mass planets. We recall that for a resonant  
capture the planets should exhibit convergent migration. Consequently, if the planets migrate inward with  
the same speed, or if the inner planet migrates even faster than the outer one, no resonant capture will occur.   
For example, if the inner planet does not open a sufficiently empty gap, it may migrate faster than a usual   
type II migration. It is also true that the inner planet's inward migration is mainly governed by the torques   
coming from the middle part of the disk, which is between the planets. (The outer disk does not really exert   
torques to the inner planet because it is separated by the wide gap opened by the outer planet.)    
On the other hand, the inward migration of the inner planet can be stopped and even reversed by a   
sufficiently massive inner disk \citep{Sandoretal2007A&A,Cridaetal2008A&A}. From our hydro simulations one   
can see that only the massive inner planet of Fit II can open a sufficiently deep gap for a slower (type II)   
migration, and could also create a massive inner disk as pushing the disk's material toward the star.   
The planets with masses from Fit I and the intermediate case (with $i=30^{\circ}$) open a shallower  
gap, the middle disk is more massive, and the inner disk is less massive for planets having   
masses from Fit II. This can be checked in Fig. \ref{avsurf_density}, where we display the azimuthally   
averaged surface density profiles of the disk for the different planetary masses. Thus, as can be seen in   
the upper panel of Fig. \ref{formation_sma_ecc}, the migration of the inner planet (of Fit II) is slow in   
the beginning and is stopped after a certain time, and right before the resonant capture (at $t\approx 2300$)  
its inward migration is even slightly reversed, which is not seen in the figure. The action of the inner disk means that 
we might expect a slow outward migration of the inner planet \citep{Cridaetal2008A&A}. However, this does not 
play any role in the resonant capture in our case, as it is the outer planet that migrates towards the inner one.
  
If the effect of the inner disk is not taken into account, the inward   
migrating outer planet always captures the inner planet into an MMR. In this case, the results   
of \citet{Kleyetal2004A&A} clearly demonstrate that the final system is either in 3:1 or in 2:1 MMR   
mainly depending on the speed of the outer planet's migration. A fast migration of the outer planet may  
result in its crossing the 3:1 MMR without capture and ending the migration in the more robust 2:1 MMR.  
  
In the case of the large planetary masses, around 2300 periods of the inner planet, the planets   
enter into the 3:1 MMR. The evolution of the semi-major axes and eccentricities during the migration   
and after the resonant capture is shown in Fig. \ref{formation_sma_ecc}. After the resonant capture, the   
eccentricities first increase and then tend to oscillate (around their mean values). However, the oscillations  
of the eccentricities remain bounded approximately in the range given by the numerical integration shown in   
Fig. \ref{ecc_sini90}. The inner disk certainly does contribute to the apparently very effective damping of   
the inner planet's eccentricity, as shown in the formation scenarios of the resonant systems HD~73526 and   
GJ~876 by \citet{Sandoretal2007A&A} and \citet{Cridaetal2008A&A}.   
  
According to our simulation, the resonant angles $\theta_i$ ($i=1,2,3$) and the corotation angle   
$\Delta\omega$ show libration, $\theta_1$ librates around $0^{\circ}$, and the remaining angles around   
$180^{\circ}$, see Figure \ref{formation_resang}.   
The dynamical behaviour of our modelled system is qualitatively the same as the observed one,  
displayed in Figure \ref{ecc_sini90}, even though in our case the libration amplitudes are somewhat smaller.  
   
Interestingly though, the behaviour of the resonant angles does not match the theoretical results  
of \citet{Beaugeetal2003ApJ} and the numerical simulations done by \citet{Kleyetal2004A&A}.   
Although the eccentricity of the inner planet exceeds the limit of $e_1\sim 0.13$, the libration of   
$\theta_3$ and $\Delta\omega$ is found to be around $180^{\circ}$.   
As already mentioned, this might be the consequence of the system not yet reaching a stationary   
solution. In what follows, we study this unexpected result through numerical integration of the   
gravitational three-body problem using properly parametrized non-conservative forces for migration.

%___________________________________________________________________________  
  
\section{Toward the stationary solutions of the 3:1 MMR in HD~60532}

In the previous section we presented a reasonably well-working formation scenario for the system  
HD~60532. The eccentricities of the giant planets and resonant angles (including the corotation angle)    
of the formed system behave very similarly to the results of numerical integration using  
initial conditions from Fit II. In this section, we intend to explain why the observed and the   
formed systems do not fit the previous studies of \citet{Beaugeetal2003ApJ} and   
\citet{Kleyetal2004A&A}.  
To do so, we performed two series of simulations; in the first one, only the outer planet is forced to 
migrate inward, while the inner planet does not feel any additional damping forces. In the second one,
the inner planet feels the (accelerating) effect of the inner disk as well.   
  
The additional non-conservative force responsible for migration and eccentrity damping  
can be parametrized by the migration rate $\dot a/a$ and the eccentricity damping rate $\dot e/e$, 
or by the corresponding $e-$folding times  $\tau_a$ and $\tau_e$ of the semimajor axis and eccentricity 
of the outer planet \citep[see][for two possible approaches]{LeeandPeale2002ApJ,Beaugeetal2006MNRAS}.   
In our simulations, we used the force corrections as suggested by \citet{LeeandPeale2002ApJ}.  
The relations between the damping rates and $e$-folding times are $\dot a/a = - 1/\tau_a$,    
and similarly for the eccentricities $\dot e/e = - 1/\tau_e$. One can define the ratio between the   
$e$-folding times $K=\tau_a/\tau_e$ (or $\dot e/e = - K |\dot a/a|$), which according to   
\citet{LeeandPeale2002ApJ} determines the final state of the system in the case of a sufficiently   
slow migration.    
  
If the inner planet is influenced by an inner disk, it may be forced to migrate outward   
and its eccentricity is also damped. This effect can also be modelled by using a   
(repelling) non-conservative force parametrized by $\dot a/a$ (having a positive sign for   
outward migration), and $\dot e/e$ for the eccentricity damping. A detailed study of the   
resulting dynamical effect of these parametrizations has been presented recently by   
\citet{Cridaetal2008A&A}.  
  
\begin{figure}  
   \centering  
   \includegraphics[width=\columnwidth]{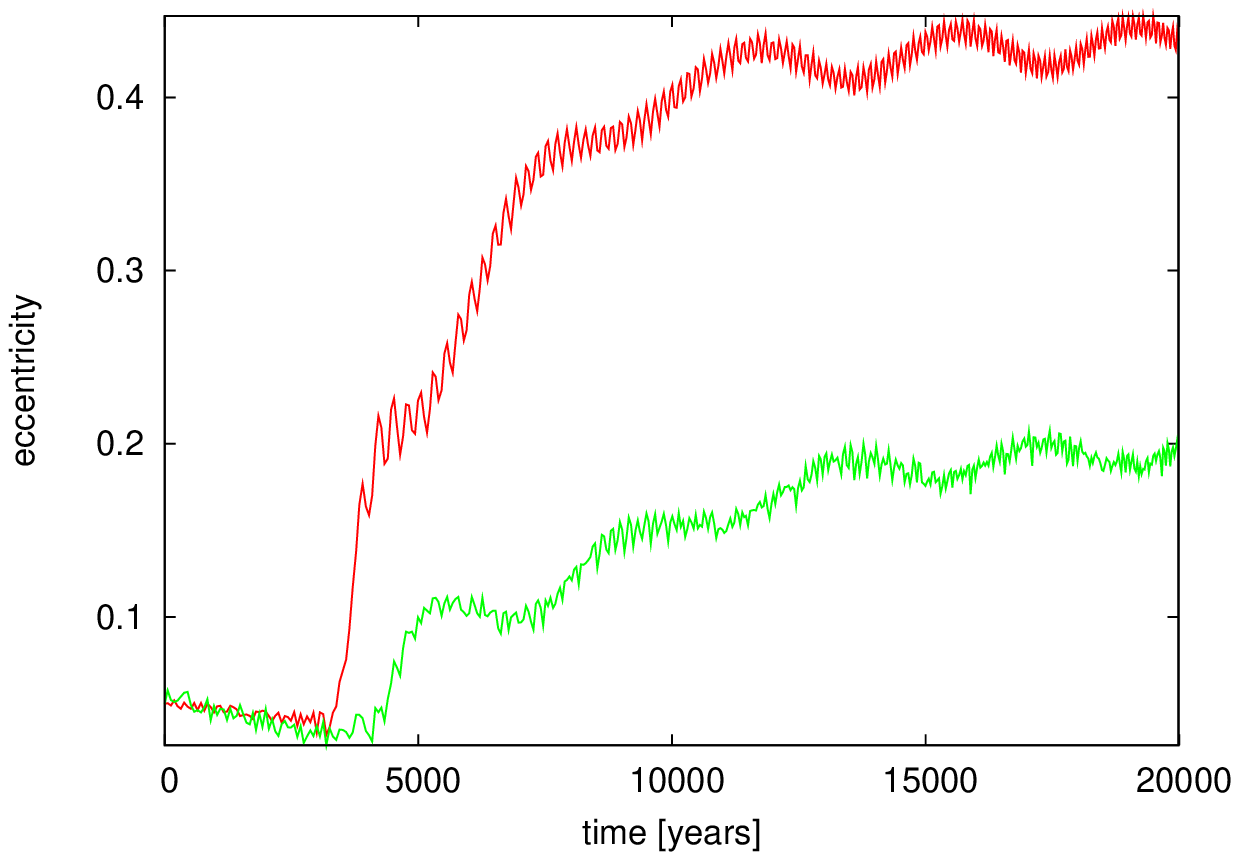}  
   \includegraphics[width=\columnwidth]{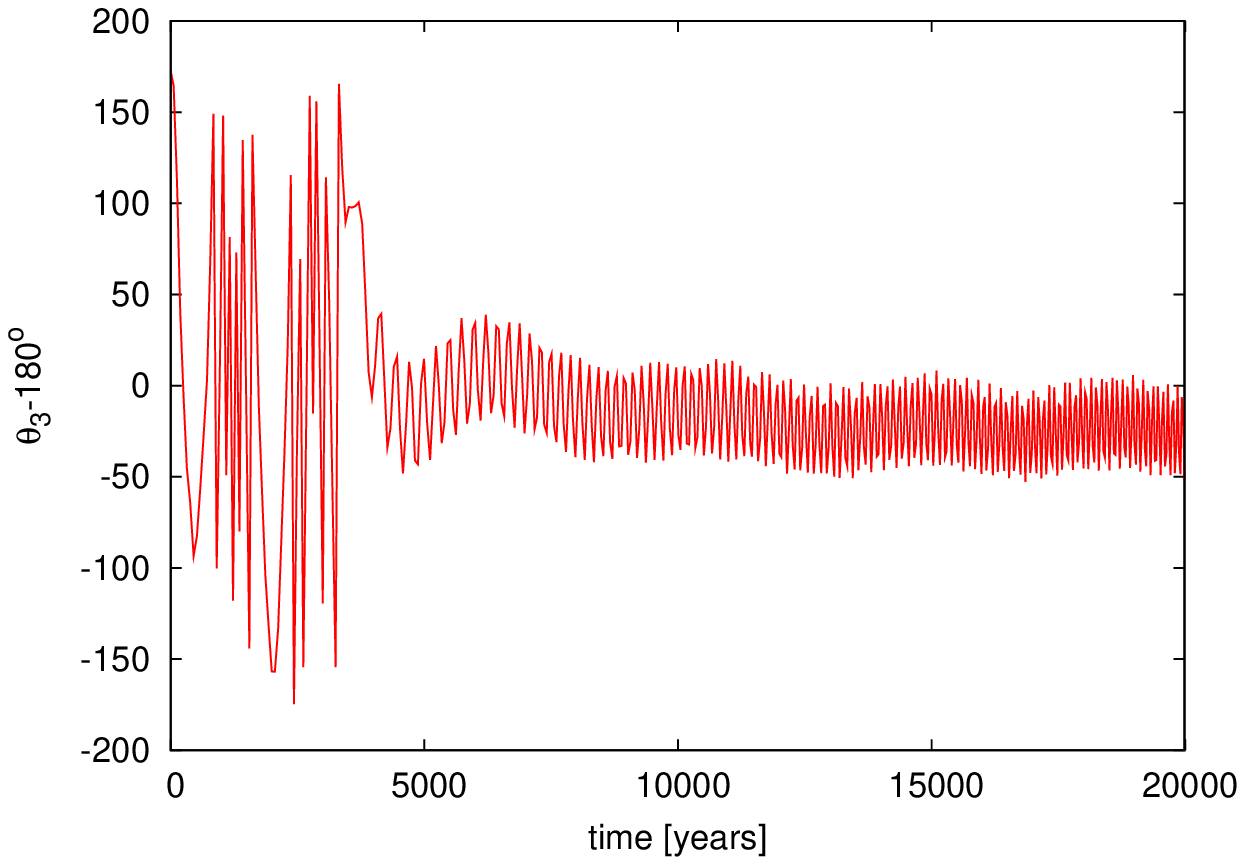}  
   \includegraphics[width=\columnwidth]{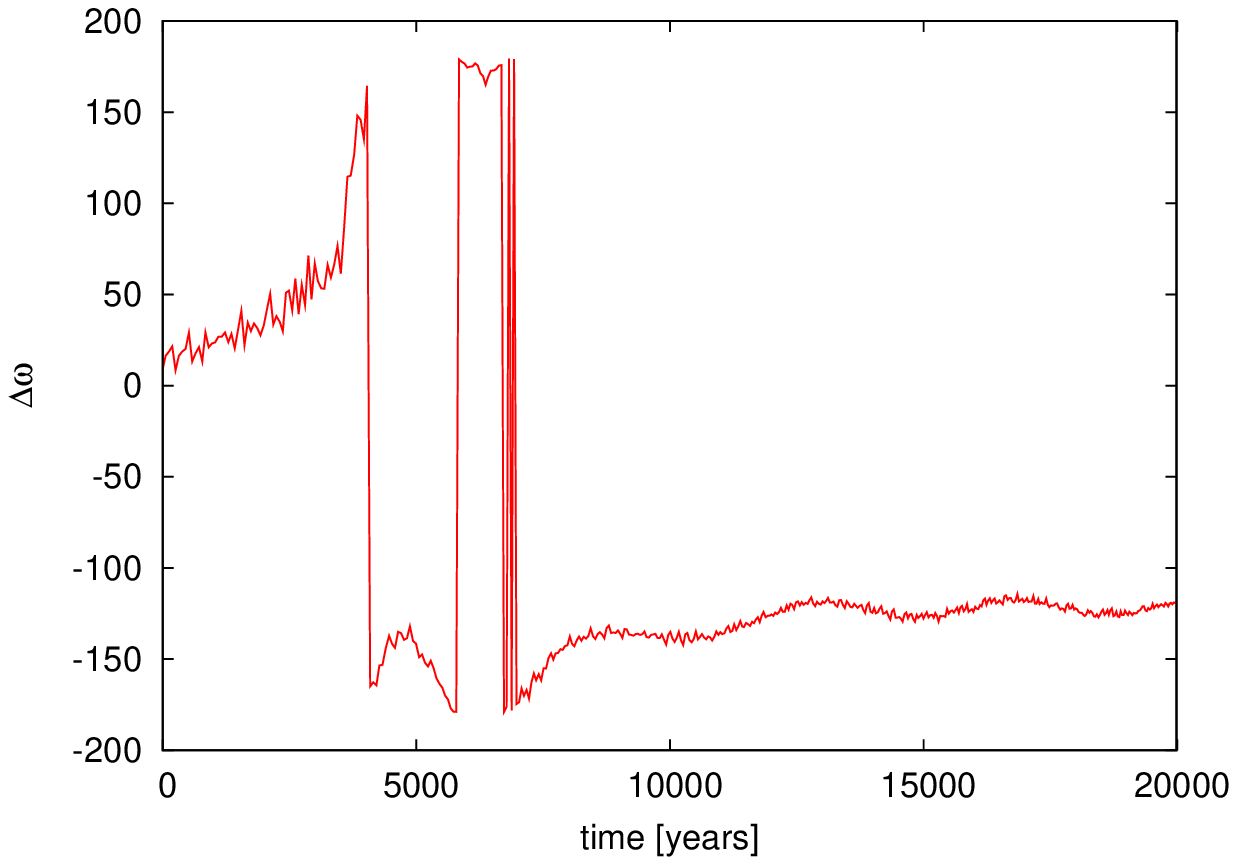}  
      \caption{Behaviour of the eccentricities (red for the inner, green for   
               the outer planet), the resonant angle $\theta_3$, and the  
	       corotation angle $\Delta\omega$ when only the outer planet  
	       migrates inward and the inner planet is not affected.}  
         \label{no_innerdamping}  
   \end{figure}

\begin{figure}  
   \centering  
   \includegraphics[width=\columnwidth]{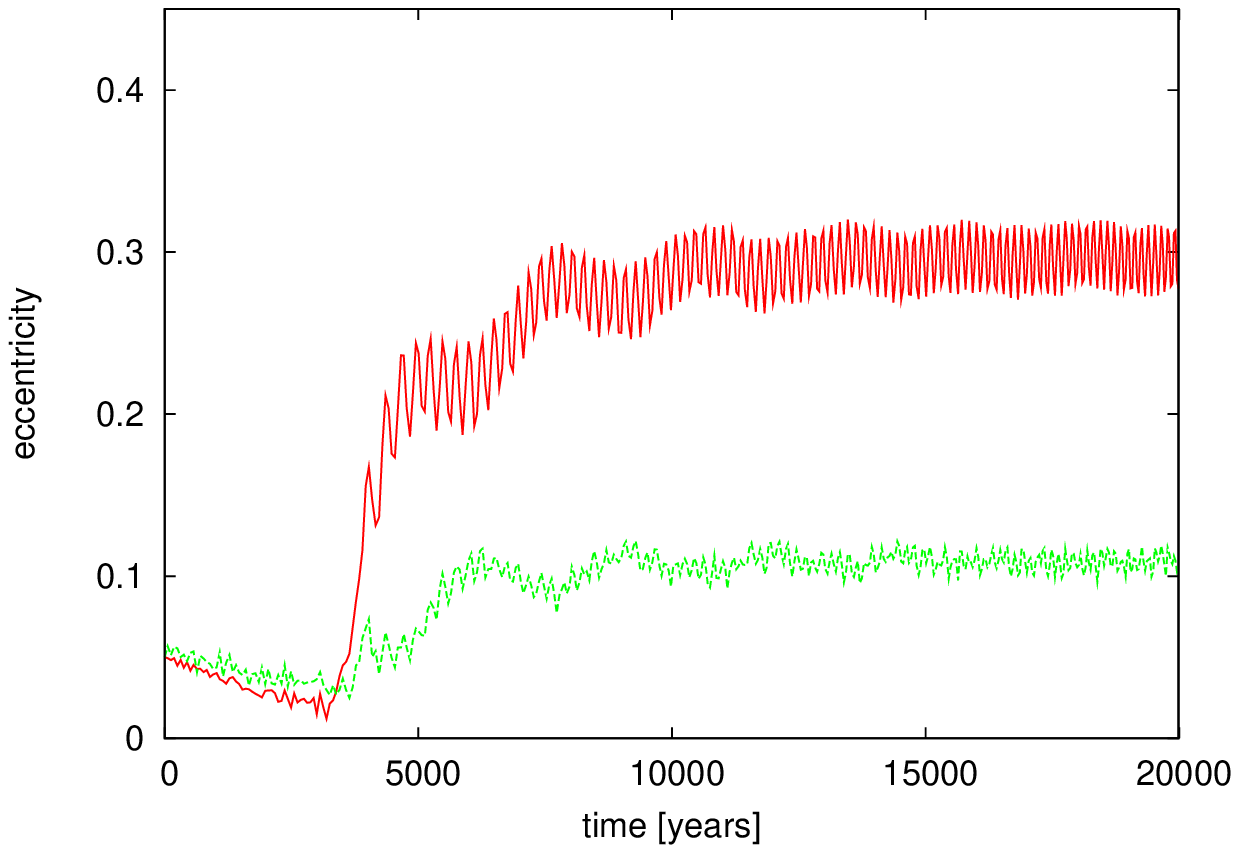}  
   \includegraphics[width=\columnwidth]{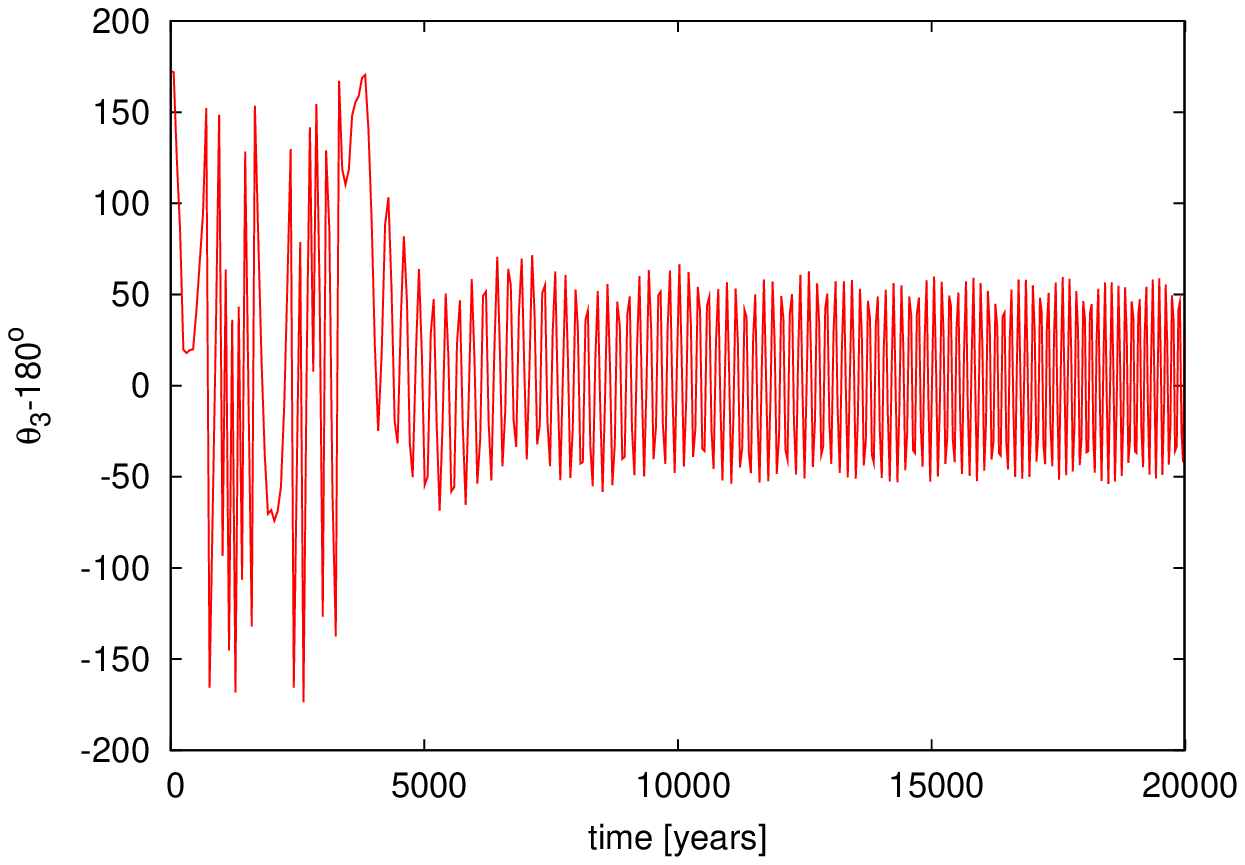}  
   \includegraphics[width=\columnwidth]{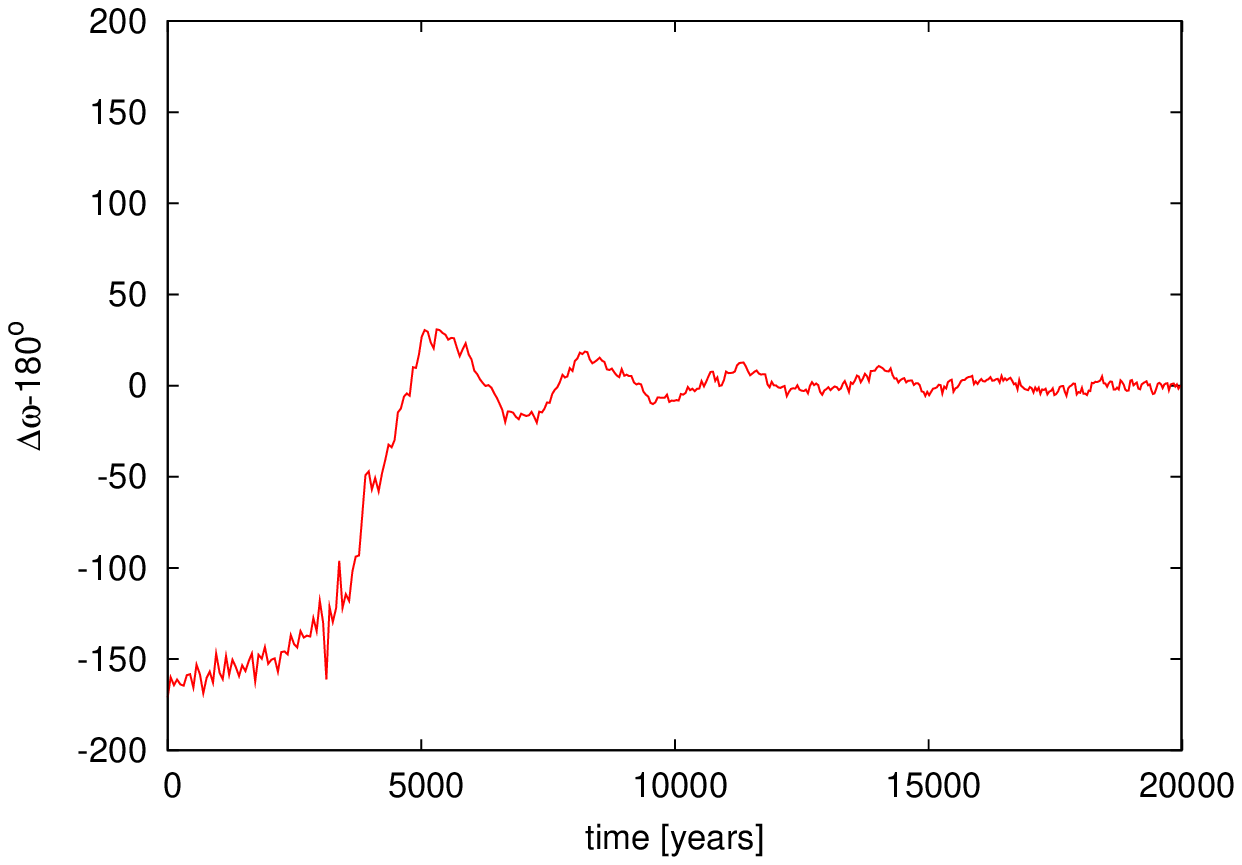}  
      \caption{Behaviour of the eccentricities (red for the inner, green for   
               the outer planet), the resonant angle $\theta_3$, and the  
	       corotation angle $\Delta\omega$ when the outer planet experiences  
	       migration and eccentricity damping, and the inner planet only the  
               eccentricity damping effect of an inner disk.}  
         \label{innerdamping}  
   \end{figure}

In the first series of simulations, only the outer planet migrated on different 
timescales $\tau_a$, while its eccentricity was damped on timescales $\tau_e$ such that their corresponding ratios 
($K=\tau_a/\tau_e$) were always between $K \sim 1...10$, which is typically found in hydrodynamical simulations 
\citep{Kleyetal2004A&A}. In the second series of runs, besides the 
migration of the outer planet, the inner planet's eccentricity was also damped, while its semi-major 
axis was not influenced.  
Typical cases of these simulations can be seen in Figs. \ref{no_innerdamping} and \ref{innerdamping}   
In these particular cases the following migration and eccentricity damping timescales were applied; for 
the outer planet  $\dot a_2/a_2=-5\times 10^{-5} \mathrm{year}^{-1}$,   
$\dot e_2/e_2=-2\times 10^{-4} \mathrm{year}^{-1}$   
(or $\tau_{a_2} = 2\times 10^4$, $\tau_{e_2} = 5\times 10^3$ years), and if the inner disk's   
influence on the inner planet was also taken into account, we used for the inner planet   
$\dot a_1/a_1=0$ (as its semi-major axis was not influenced) and   
$\dot e_1/e_1=-2\times 10^{-4} \mathrm{year}^{-1}$.  
   
Comparing Figs. \ref{no_innerdamping} and \ref{innerdamping}, it can be seen immediately that,  
in the first case when the inner planet was unaffected, the resonant angles behave according  
to the theoretical predictions of \citet{Beaugeetal2003ApJ} (asymmetric libration of the resonant  
angles), while in the second case the libration of the resonant angles is around $180^{\circ}$  
and $0^{\circ}$, respectively. The inclusion of the inner disk helps to keep the inner  
planet's eccentricity at lower values close to those coming from the observations. In Fig.   
\ref{no_innerdamping} the eccentricity of the inner planet tends to 0.5, while in Fig.  
\ref{innerdamping} it will show only small amplitude oscillations around the value 0.3.  
  
Finally, we can conclude that the effect of an inner disk in the  
three-body simulations (with dissipative forces) alters the final state of the resonant   
configuration (here a 3:1 MMR), keeping the resonant angles far from their mean libration, 
thus impeding the system for reaching the equilibrium solutions suggested by \citet{Beaugeetal2003ApJ}.    
In both cases the systems will remain in their final configrations, when the damping effect of the 
disk is reduced and eventually non existent.
  
%-------------------------------------------------------------  
%--------------------------------------------------------------  
   
%-------------------------------------------------------------  
%--------------------------------------------------------------   

%____________________________________________________________________________  
\section{Summary}  

After the discovery of the system HD 60532 \citep{Desortetal2008A&A},
the thorough dynamical analysis of \citet{LaskarandCorreia2009A&A} has reliably established 
the first resonant system that contains two giant planets in a higher order mean motion resonance, 
here 3:1. Until now, only one system, 55 Cancri, has been proposed as a candidate for a system  
where two giant planets might be in 3:1 MMR. On the other hand, for 55 Cancri,   
the existence of the resonant solution could not be confirmed by \citet{Naefetal2004A&A}, and a new  
non-resonant solution was later found by \citet{Fischeretal2008ApJ}. Thus according to our present knowledge,   
HD~60532 is the only known system containing giant planets in the 3:1 MMR.  
    
Recently, \citet{LaskarandCorreia2009A&A} has improved the orbital solution found by \citet{Desortetal2008A&A}.   
In their new fit that assumes a relatively small inclination between the common orbital plane of the planets  
and the tangent plane of the sky, $i=20^{\circ}$, the giant planets have quite high masses:   
$m_1=3.15 M_J$ and $m_2=7.46 M_J$.   
  
The main path to forming a resonant planetary system is thought to be by convergent migration of planets.  
For an alternative scenario, formation through a scattering process has been proposed by \citet{2008ApJ...687L.107R}.  
Migration is the the result of dissipative forces originating in an ambient accretion disk,  
which act along with the mutual gravitational forces between the planets and the central star. 
Being a dissipative process, a sufficiently long-lasting migration process brings the system close to a stationary 
solution, which corresponds to a periodic solution of the system.      
The stationary (minimum energy) solutions in the systems yielding 3:1 MMR have been studied by a Hamiltonian approach  
by \citet{Beaugeetal2003ApJ}. It has been found that, depending on the ratio of the eccentricities,   
the resonant angles can exhibit antisymmetric (symmetric anti-aligned) or asymmetric libration. These results have  
also been confirmed by numerical simulations of \citet{Kleyetal2004A&A}.   
  
In a first step, we integrated the system HD~60532 numerically using the (observed) initial conditions   
calculated from Fits I and II of \citet{LaskarandCorreia2009A&A}.   
We found that all resonant angles librate and that the system is in apsidal corotation.  
Since $\Delta \omega$ oscillates around $180^{\circ}$, observationally the system lies in an antisymmetric configuration. 
However, for the observed mean eccentricities, the Hamiltonian approach would suggest an asymmetric configuration.  
  
To model the formation of the system HD~60532 and find an explanation for this special antisymmetric  
configuration, we assumed that the system formed through a planet-disk interaction process.  
We performed a series of fully hydrodynamical simulations with low,   
intermediate, and high planetary masses $m_i/\sin i$, assuming $i=90^\circ$, $30^\circ$, and $20^\circ$,   
respectively. Since the eccentricity of the inner planet  
oscillates around a moderate mean value $e_1\sim 0.3$, we assumed that during the migration process  
an inner disk (between the inner planet and the central star) was present, providing an efficient  
damping mechanism on the inner planet's eccentricity.   
We found that the convergent migration through planet-disk interaction, which takes the planets into  
the 3:1 MMR only occurred for the highest planetary masses. The dynamical  
behaviour of the resulting resonant planetary system is then indeed very similar to the one   
obtained from the radial velocity observations. 
Through our full hydrodynamical simulations, we support the small inclination
$i=20^{\circ}$ of the system as suggested by \citet{LaskarandCorreia2009A&A}.
  
To understand why the system does not appear to be in a miminum energy configuration to the 3:1 MMR,  
we performed a series of dedicated 3-body simulations with additonal forces emulating the effects of  
the protoplanetary disk. In particular, we studied situations with and without the influence of an inner  
disk.  We found that it is exactly the effect of the inner disk that distiguishes between an antisymmetric  
or asymmetric final configuration. Here, its presence is responsible for the system not reaching the minimum 
energy asymmetric configuration. Nevertheless, we point out that the solutions obtained in the antisymmetric 
state are indeed stationary solutions, which are stable when the effects of the disk are reduced. 
  
In related earlier works, we had already established that the observed smallness of the mean eccentricities in  
resonant planetary systems can also be attributed to the effect of an inner disk   
\citep{Sandoretal2007A&A,Cridaetal2008A&A}.   
We conclude that an inner disk during the migration process has directly observable   
consequences on the post-formation dynamical behaviour of resonant planetary systems in  
mean motion resonances.   
  
\begin{acknowledgements}  
 This work was partly supported by the German Research Foundation (DFG) under grant KL~650/11 
 within the collaborative research group {\it The formation of planets} (FOR~759).  
\end{acknowledgements}  
  
\bibliographystyle{aa}  
  
\bibliography{sk2010bib}  
  
\end{document}